\documentclass[prd,showpacs,preprint,nofootinbib]{revtex4}
\usepackage{graphicx,color,amsmath}
\usepackage{bm}
\usepackage{slashed}
\allowdisplaybreaks
\newcommand{\D}{{\rm{d}}}

\begin{document}
\title{Infrared scalar one-loop three point integrals in loop regularization}
\author{Jin Zhang\footnote{jinzhang@yxnu.edu.cn}}
\affiliation{School of Physics and Engineering, Yuxi Normal University,
Yuxi, Yunnan, 653100, P. R. China}


\begin{abstract}
The infrared divergent scalar three-point integrals are
evaluated by the loop regularization method. Three kinds
of infrared divergent integrals, i.e., massless triangle diagram,
triangle diagrams with one and two massive internal lines,
are systematically evaluated by loop regularization, analytic
results are obtained. According the
method, the infrared divergences
are regulated by the so-called sliding scale $\mu_{s}$ which plays the
role of infrared cutoff. The amplitudes obtained through loop regularization
depend on $\mu_{s}$ such that we may extract different contribution
by varying $\mu_{s}$. Some general results for evaluation of
scale one-loop triangle diagram are also derived.
\end{abstract}


\maketitle
\newpage

\section{introduction}
The evaluation of amplitudes of the three-point(triangle)
Feynman diagrams holds a prominent position both in renormalization or extracting
infrared stable cross section at one-loop level within the
framework of the Standard Model(SM) which can be read from any standard textbook
on Quantum Field Theory(QFT)\cite{Peskin:1995ev, Weinberg:1995mt}.
It is known that there are processes generated only
through triangle diagram, for instance,
producing the Higgs bosons by gluon fusion, Higgs decaying to two gluons or
two photons\cite{Ellis:1975ap, Rizzo:1979mf, Djouadi:2005gi, Workman:2022ynf}.
If there is massless field like photon in QED or
gluon in QCD, we need to deal with the infrared divergence in evaluation of the one-loop triangle diagrams.
Furthermore, if the massless field couples to other massless fields or itself,
another infrared divergence called mass singularity(or collinear divergence) will emerge.
In line with the standard procedure in evaluating Feynman diagrams,
an appropriate regularization scheme must be employed such that the
primitive divergent integrals can be well defined mathematically,
then the infrared divergent parts and the infrared stable parts can be
separated definitely. The most popular regularization scheme
is the dimensional regularization\cite{tHooft:1978jhc, RevModPhys.47.849} by which we move from the
four-dimensional space-time to an arbitrary $d$-dimension,
when the integrals over intermediate momentum and Feynman parameters are completed,
the divergences will appear as the pole terms $1/\varepsilon$ with $\varepsilon=d-4$.

If there is only infrared divergence in the evaluation of the amplitudes,
an alternative way to extract the divergence is the so-called loop
regularization\cite{Wu:2002xa, Wu:2003dx}. According to this method, we introduce a sliding
scale $\mu_{s}$ to regulate the infrared divergence,
the meaning of $\mu_{s}$ is that it amounts to
assign a mass $\mu_{s}$ uniformly to each internal line.
This approach has been successfully applied in evaluating the CP violation
in hadronic weak decays of $B$ mesons\cite{Su:2010vt, Su:2011eq}
and in evaluation of the amplitude
of Higgs decaying into two photons\cite{Huang:2011yf} which is helpful in clarifying
the gauge invariance issue\cite{Shifman:2011ri, Marciano:2011gm}
raised in\cite{Gastmans:2011ks, Gastmans:2011wh}.
Recently, the loop regularization scheme is also exploited
in evaluation of infrared divergent one-loop four points amplitude
with one massless vertex\cite{Zhang:2021skq}, it turns out that this method works well.

We notice that although loop regularization finds its use in
hadronic weak decays, there is still lack of a full
study on the infrared divergent one-loop triangle diagram.
On the other hand, in considering the significant role played by the one-loop
triangle diagram in describing some decaying process, in this paper
the whole infrared divergent one-loop triangle diagrams will be evaluated
by the loop regularization scheme. We hope the work in this paper
can shed some light on the one-loop triangle evaluation and decays mediated
by triangle at lowest order of perturbation theory.

Before starting the evaluation, the following comments are in order.

(\rm{$\romannumeral 1$}) It is worth emphasizing the difference between
massive gluon\footnote{A thorough application of massive gluon scheme to deal with
one-loop triangle diagrams and other processes in QCD as well as comparison with dimensional
regularization, one may refer to\cite{RDField}.}
method and loop regularization since in this two scheme,
In the massive gluon scheme, in order to regulate the infrared divergence,
one assumes that the only internal gluon has some fictitious mass $m_{g}$ and
keep the mass of fermion propagators unchanged. While in loop regularization,
all the internal lines, both the fermions and bosons, acquire the same mass $\mu_{s}$.

\begin{figure}
\begin{center}
\includegraphics[scale=0.45]{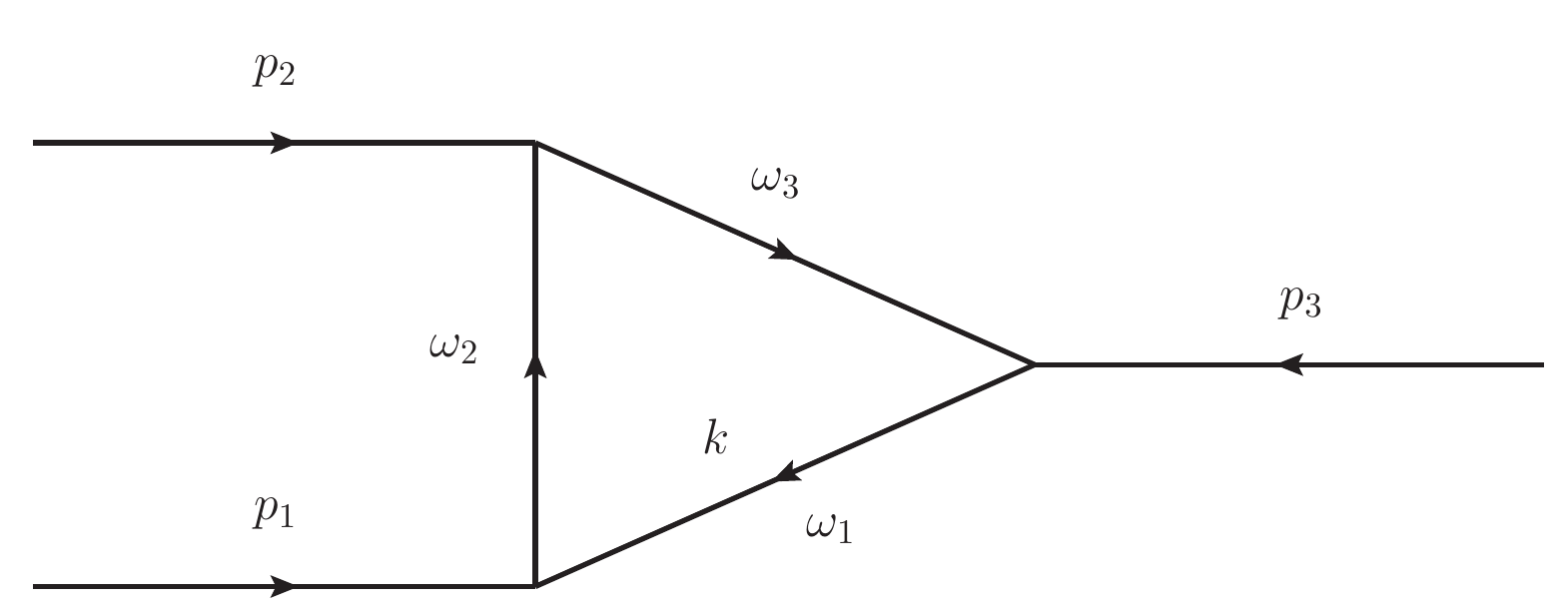}
\caption{The most general scalar one-loop 3-point integral.
As usual all the external momenta are
inward. The $\omega_{i}(i=1, 2, 3)$
denote the mass of each internal lines, all the external lines are
mass-shell.}\label{scalarthreepoint}
\end{center}
\end{figure}

(\rm{$\romannumeral 2$}) The evaluation in this paper is
inspired by producing Higgs boson thorough gluon fusion,
hence in order to be consistent with the real physical picture, if
there are more than one massive internal lineswe assume
that these lines have equal mass. In addition, we assume all the external
lines are on mass-shell
\begin{equation}
p_{1}^{2}=m_{1}^{2}, \quad\quad p_{2}^{2}=m_{2}^{2},
\quad\quad p_{3}^{2}=m_{3}^{2}.  \label{externalonmassshell}
\end{equation}
and setting $m_{3}>m_{1}, m_{2}$. The most general triangle diagram is depicted
in fig.\ref{scalarthreepoint}.

(\rm{$\romannumeral 3$}) We do not so ambitious as some
existing work\cite{Ellis:2007qk, tHooft:1978jhc, vanHameren:2010cp} that including
both the analytic evaluation of one-loop triangle diagrams and applications to specific
processes, in this paper we content ourselves with getting analytic results of infrared divergent one-loop
triangle diagrams by loop regularization but without phenomenological applications.

The rest of the paper is structured as follows.
In section $\rm{\uppercase\expandafter{\romannumeral 2}}$ we first
briefly discuss the general form of one-loop triangle integral, then by making use the
loop regularization, the divergent integrals are evaluated case-by-case, infrared divergent parts and the
infrared stable are obtained. A brief discussion and summary is presented in section
$\rm{\uppercase\expandafter{\romannumeral 3}}$.
Some necessary materials are contained in the appendix.

\section{the formulas}

The most general scalar one loop three-point integral is depicted
in Fig.\ref{scalarthreepoint}, the amplitude is
\begin{equation}
I=\int\frac{\D^{4} k}{(2\pi)^{4}}
\frac{1}{(k^{2}-\omega_{1}^{2}-i\varepsilon)
[(k+p_{1})^{2}-\omega_{2}^{2}-i\varepsilon]
[(k+p_{1}+p_{2})^{2}-\omega_{3}^{2}-i\varepsilon]}, \label{mostgeneralscalarthree}
\end{equation}
By using Feynman parameterization, the primitive integral can be written as
\begin{equation}
I=\int\frac{\D^{4} k}{(2\pi)^{4}}
\int\D x\D y\D z\delta(1-x-y-z)\frac{2}{D^{3}}, \label{feynmanparameterization}
\end{equation}
the denominator in Eq.(\ref{feynmanparameterization}) is
\begin{equation}
D=x(k^{2}-\omega_{1}^{2})+y[(k+p_{1})^{2}-\omega_{2}^{2}-i\varepsilon]
+z[(k+p_{1}+p_{2})^{2}-\omega_{3}^{2}-i\varepsilon]. \label{denamoinatorafterfeynamnpara}
\end{equation}
where $\varepsilon\rightarrow 0^{+}$.
If all the internal masses are nonzero, there will be no
infrared divergence in Eq.(\ref{feynmanparameterization}),
To make its appearance of infrared divergence,
one or more internal lines should be massless. In what follows
the one-loop triangle with one-, two- and three massless
internal lines will be evaluated, the infrared divergence
will be extracted by loop regularization.According to the method,
the loop momentum $k$ transforms in the following manner
\begin{eqnarray}
&&k^{2}\rightarrow [k^{2}]_{l}=k^{2}-M_{l}^{2},\nonumber\\
&&\int\frac{\D^{4}k}{(2\pi)^{4}}
\rightarrow\int\Big[\frac{\D^{4}k}{(2\pi)^{4}}\Big]_{l}
=\lim_{N, M_{i}^{2}
\rightarrow\infty}\sum_{l=0}^{N}c^{N}_{l}\int\frac{d^{4}k}{(2\pi)^{4}}, \label{loopregul}
\end{eqnarray}
which is constrained by
\begin{eqnarray}
\lim_{N, M_{i}^{2}\rightarrow\infty}\sum_{l=0}^{N}c^{N}_{l}(M_{l}^{2})^{2}=0,\quad
c_{0}^{N}=0 \quad (i=0,1,...,N\,\,\text{and}\,\, n=0,1,..).\label{loopconstrain}
\end{eqnarray}
From Eq.(\ref{loopconstrain}) the coefficients $c_{l}^{N}$ can be worked out
\begin{equation}
c_{l}^{N}=(-1)^{l}\frac{N!}{l!(N-l)!},\nonumber
\end{equation}
the regulator mass is given by
\begin{equation}
M^{2}_{l}=\mu^{2}_{s}+lM^{2}_{R},\label{regulatormass}
\end{equation}
Then it leads to the desired integration form over $k$
\begin{eqnarray}
&&k^{2}\rightarrow k^{2}-\mu^{2}_{s}-lM^{2}_{R},\nonumber\\
&&\int\frac{\D^{4}k}{(2\pi)^{4}}\rightarrow
\lim_{N, M_{R}^{2}\rightarrow\infty}\sum_{l=0}^{N}(-1)^{l}
\frac{N!}{l!(N-l)!}\int\frac{\D^{4}k}{(2\pi)^{4}}.\label{loopmomentum}
\end{eqnarray}
If there were only infrared divergence, when the integration
over loop momentum is completed, terms involving $M_{R}$ will vanish
after taking the limit, thus in this case it is equivalent to
introduce a characteristic scale $\mu_{s}$ in the amplitudes.
After integration over the intermediate momentum
in Eq.(\ref{feynmanparameterization}) is completed, getting
\begin{eqnarray}
I= \frac{-i}{(4\pi)^{2}}\int\D x\D y\D z
\delta(1-x-y-z)\frac{1}{\Delta(x,y,z)}, \label{intermediateintegrated}
\end{eqnarray}
where
\begin{eqnarray}
\Delta(x,y,z)&=&y^{2}p_{1}^{2}+z^{2}p_{3}^{2}
-yz(p_{2}^{2}-p_{1}^{2}-p_{3}^{2})\nonumber\\
&+&x\omega_{1}^{2}-y(p_{1}^{2}-\omega_{2}^{2})
-z(p_{3}^{2}-\omega_{3}^{2})
+\mu_{s}^{2}-i\varepsilon, \label{feynmanparadenominator}
\end{eqnarray}
Sine the integral in Eq.(\ref{intermediateintegrated}) is restricted by the $\delta$ function,
we can see that in loop regularization, it amounts to assign an extra mass $\mu_{s}$ to each
propagator. The integration over $z$ of Eq.(\ref{intermediateintegrated}) is trivial,
the result reads
\begin{equation}
I=\frac{-i}{(4\pi)^{2}}\int_{0}^{1}\D x
\int_{0}^{1-x}\D y\frac{1}{\Delta(x,y,1-x-y)}.  \label{zintegrated}
\end{equation}
In the forthcoming sections the pre-factor $-i/(4\pi)$ will be suppressed for brevity.

\subsection{three massless triangle}

\begin{figure}
\begin{center}
\includegraphics[scale=0.75]{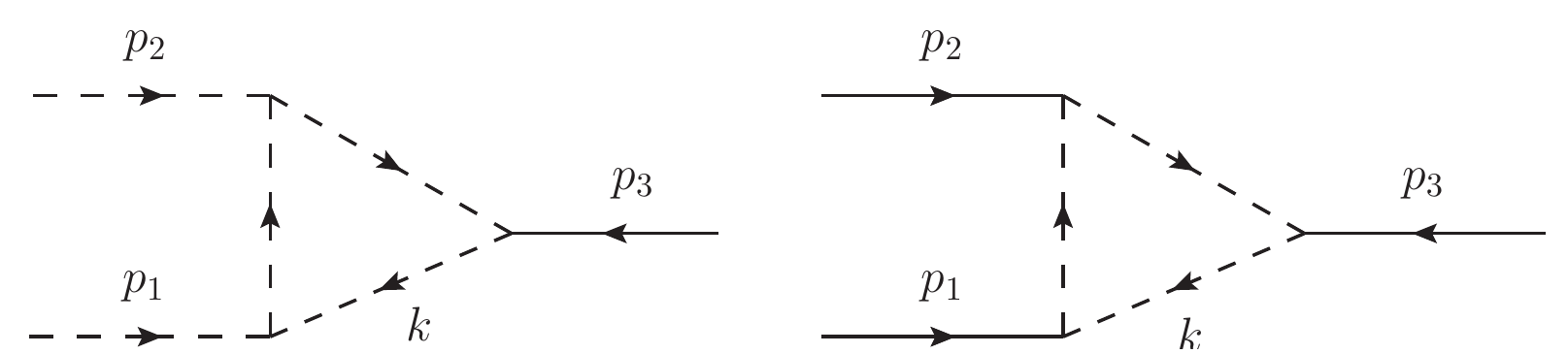}
\caption{Massless triangle with one-massive(left)
and three-massive external lines(right). The solid lines and dashed
lines denote massive and massless particles, respectively,}\label{masslesstriangle}
\end{center}
\end{figure}

(\rm{$\romannumeral 1$}) $\omega_{1}=\omega_{2}=\omega_{3}=0,
p_{1}^{2}=p_{2}^{2}=0, p_{3}^{2}=m^{2}$

The diagrams is depicted on the left of fig.\ref{masslesstriangle},
according to Eq.(\ref{intermediateintegrated}) we can write the amplitude
\begin{eqnarray}
I&=&\int_{0}^{1}\D x\int_{0}^{1-x}\D y
\frac{1}{x^{2}m^{2}+xym^{2}-xm^{2}
+\mu_{s}^{2}-i\varepsilon}\nonumber\\
&=&\frac{1}{m^{2}}\int_{0}^{1}\D x
\frac{\ln\big(\frac{m^{2}}{\mu_{s}^{2}}x^{2}
-\frac{m^{2}}{\mu_{s}^{2}}x+1-i\varepsilon\big)}{x}, \label{caseoneayintegrated}
\end{eqnarray}
By employing Eq.(\ref{finalresultofspecialcase})
and Eq.(\ref{finalresultaequaltob}), we obtain
\begin{equation}
I=\frac{1}{m^{2}}\big[\arcsin(\frac{m}{2\mu_{s}})\big]^{2},\,
\quad\quad\quad\quad 0<m<2\mu_{s}  \label{masslesstriangleresultone}
\end{equation}
and
\begin{equation}
I=\frac{1}{2m^{2}}\Big(\pi+i\ln\frac{1+\sqrt{1-\frac{4\mu_{s}^{2}}{m^{2}}}}
{1-\sqrt{1-\frac{4\mu_{s}^{2}}{m^{2}}}}\Big)^{2}. \quad m>2\mu_{s} \label{masslesstriangleresulttwo}
\end{equation}
From the above results it is obvious that $\mu_{s}$ plays the role of
infrared cutoff.

(\rm{$\romannumeral 2$})\quad $ \omega_{1}=\omega_{2}=\omega_{3}=0,
p_{1}^{2}=m_{1}^{2}, p_{2}^{2}=m_{2}^{2}, p_{3}^{2}=m_{3}^{2}, m_{3}>m_{1},m_{2}$.

The diagram is depicted on the right of fig.\ref{masslesstriangle}.
The amplitude is
\begin{eqnarray}
I=\int_{0}^{1}\D x\int_{0}^{1-x}\D y\frac{1}{\Delta(x,y)},   \label{caseatwofirst}
\end{eqnarray}
where
\begin{equation}
\Delta(x,y)=x^{2}m_{3}^{2}+y^{2}m_{2}^{2}
+xy(m_{2}^{2}+m_{3}^{2}-m_{1}^{2})
-xm_{3}^{2}-ym_{2}^{2}+\mu_{s}^{2}-i\varepsilon,   \label{denomenatorcaseatwo}
\end{equation}
To proceed we make variable transformation
\begin{equation}
x=1-x',\quad y=x'-y',   \label{caseatwofristsub}
\end{equation}
with the Jacobian
\begin{equation}
\frac{\partial (x,y)}{\partial(x',y')}=1  \label{jacobianmatrix}
\end{equation}
Then Eq.(\ref{caseatwofirst}) takes the following form
\begin{equation}
I=\int_{0}^{1}\D x'\int_{0}^{x'}\D y'
\frac{1}{W_{1}(x',y')},     \label{caseatwowdenomenator}
\end{equation}
the denominator $W(x',y')$ is given by
\begin{equation}
W_{1}(x',y')=x'^{2}m_{1}^{2}+y'm_{2}^{2}+x'y'(m_{3}^{2}-m_{1}^{2}-m_{2}^{2})-x'm_{1}^{2}
+y'(m_{1}^{2}-m_{2}^{2})+\mu_{s}^{2}-i\varepsilon.  \label{obviouswexpression}
\end{equation}
Since $W(x',y')$ is quadratic homogeneous polynomial in $x'$ and $y'$,
it is convenient to perform the so-called \emph{Euler-shift} on $y'$
\begin{equation}
y'=\rho+\alpha x'
\end{equation}
This leads to
\begin{equation}
I=\big[\int_{-\alpha}^{0}\D \rho\int^{1}_{-\rho/\alpha}\D x'+
\int_{0}^{1-\alpha}\D \rho\int_{\rho/(1-\alpha)}^{1}\D x'\big]
\frac{1}{W_{2}(x',\rho)}, \label{primedvaribletransform}
\end{equation}
the function $W_{2}(x',\rho)$ is defined as
\begin{eqnarray}
W_{2}(x',\rho)&=&[m_{2}^{2}\alpha^{2}+(m_{3}^{2}-m_{1}^{2}
-m_{2}^{2})\alpha+m_{1}^{2}]x'^{2}+m_{2}^{2}\rho^{2}\nonumber\\
&+&[(1+2\alpha)m_{3}^{2}-m_{2}^{2}-m_{1}^{2}]x'\rho
+[\alpha(m_{1}^{2}-m_{3}^{2})-m_{1}^{2}]x'\nonumber\\
&+&(m_{1}^{2}-m_{3}^{2})\rho+\mu_{s}^{2}-i\varepsilon, \label{wfunctionofprimedvar}
\end{eqnarray}
The the parameter $\alpha$ is chosen to obey the condition
\begin{equation}
m_{2}^{2}\alpha^{2}
+(m_{3}^{2}-m_{1}^{2}-m_{2}^{2})\alpha+m_{1}^{2}=0, \label{conatrainingonalpha}
\end{equation}
From Eq.(\ref{conatrainingonalpha}) the parameter $\alpha$ can be solved out
\begin{equation}
\alpha=\frac{1}{2m_{2}^{2}}\big[(m_{1}^{2}+m_{2}^{2}-m_{3}^{2})
\pm\lambda^{1/2}(m_{3}^{2},m_{1}^{2},m_{2}^{2})\big], \label{lamdbafunction}
\end{equation}
where $\lambda(x,y,z)$ is the K$\ddot{a}$llen function
\begin{equation}
\lambda(x,y,z)=x^{2}+y^{2}+z^{2}-2xy-2yz-2xz, \label{kallenfunction}
\end{equation}
Then we get
\begin{equation}
I=\big[\int_{-\alpha}^{0}\D \rho\int^{1}_{-\rho/\alpha}\D x'+
\int_{0}^{1-\alpha}\D \rho\int_{\rho/(1-\alpha)}^{1}\D x'\big]
\frac{1}{W_{3}(x',\rho)},  \label{linearxprimeddenomenator}
\end{equation}
the denominator is linear in $x'$
\begin{eqnarray}
W_{3}(x',\rho)&=&m_{2}^{2}\rho^{2}
+[(1+2\alpha)m_{3}^{2}-m_{2}^{2}-m_{1}^{2}]x'\rho
+[\alpha(m_{1}^{2}-m_{3}^{2})-m_{1}^{2}]x'\nonumber\\
&+&(m_{1}^{2}-m_{3}^{2})\rho+\mu_{s}^{2}-i\varepsilon. \label{wthrefunction}
\end{eqnarray}
Now the integral over $x'$ in Eq.(\ref{linearxprimeddenomenator}) is easy to calculate
\begin{eqnarray}
I&=&\int_{-\alpha}^{0}\D \rho\frac{1}{\beta_{0}+\beta_{1}\rho }
\big\{\ln\big[m^{2}_{2}\rho^{2}+(2\alpha-1)m_{2}^{2}\rho
+(\alpha-1)m_{1}^{2}-\alpha m_{3}^{2}+\mu_{s}^{2}-i\varepsilon\big]\nonumber\\
&-&\ln\big[\frac{m_{1}^{2}+(1-\alpha)m_{2}^{2}-m_{3}^{2}}{\alpha}\rho^{2}
+\frac{m_{1}^{2}}{\alpha}+\mu_{s}^{2}-i\varepsilon\big]\big\}\nonumber\\
&+&\int_{0}^{1-\alpha}\D \rho\frac{1}{\beta_{1}\rho+\beta_{0}}
\big\{\ln\big[m_{2}\rho^{2}+(2\alpha-1)m_{2}^{2}\rho
+(\alpha-1)m_{1}^{2}-m_{3}^{2}+\mu_{s}^{2}-i\varepsilon\big]\nonumber\\
&-&\ln\big(\frac{m_{1}^{2}-m_{3}^{2}-\alpha m_{2}^{2}}{\alpha-1}\rho^{2}
+\frac{m_{3}^{2}}{\alpha-1}\rho+\mu_{s}^{2}-i\varepsilon\big)\big\}, \label{xprimeintegrated}
\end{eqnarray}
where
\begin{equation}
\beta_{0}=(\alpha-1)m_{1}^{2}-\alpha m_{3}^{2},\quad
\beta_{1}=(1+2\alpha)m_{3}^{2}-m_{1}^{2}-m_{2}^{2}. \label{betaoneandbetazero}
\end{equation}
To regulate the lower limit of the first integral and the upper limit of the second integral
in Eq.(\ref{xprimeintegrated}), we make the following transformation
\begin{equation}
\rho=-\alpha\xi, \label{varrhotoxifirst}
\end{equation}
in the first integral and
\begin{equation}
\rho=(1-\alpha)\xi,  \label{rhotoxitwice}
\end{equation}
in the second integral. Hence Eq.(\ref{xprimeintegrated}) is cast into
\begin{eqnarray}
I&=&\alpha\int_{0}^{1}\D \xi
\frac{1}{\beta_{0}-\beta_{1}\alpha\xi}
\big[\ln\big(a_{1}\xi^{2}
+b_{1}\xi+c_{1}-i\varepsilon\big)
-\ln\big(a_{2}\xi^{2}+b_{2}\xi
+c_{2}-i\varepsilon\big)\big]\nonumber\\
&+&(1-\alpha)\int_{0}^{1}\D \xi
\frac{1}{\beta_{0}+(1-\alpha)\beta_{1}\xi}
\big[\ln(a_{3}\xi^{2}
+b_{3}\xi+c_{3}-i\varepsilon)\nonumber\\
&-&\ln(a_{4}\xi^{2}+b_{4}\xi+c_{4}-i\varepsilon)\big], \label{integralIonesimplified}
\end{eqnarray}
where the coefficients are
\begin{eqnarray}
a_{1}&=&\alpha^{2}m_{2}^{2},\quad
b_{1}=-\alpha(2\alpha-1)m_{2}^{2},\quad
c_{1}=(\alpha-1)m_{1}^{2}-\alpha m_{3}^{2}+\mu_{s}^{2},\nonumber\\
a_{2}&=&\alpha[m_{1}^{2}+(1-\alpha)m_{2}^{2}-m_{3}^{2}],\quad
b_{2}=-m_{1}^{2},\quad c_{2}=\mu_{s}^{2},\nonumber\\
a_{3}&=&(1-\alpha)^{2}m_{2}^{2},\quad
b_{3}=(1-\alpha)(2\alpha-1)m_{2}^{2},\quad
c_{3}=(\alpha-1)m_{1}^{2}
-\alpha m_{3}^{2}+\mu_{s}^{2},\nonumber\\
a_{4}&=&(1-\alpha)(\alpha m_{2}^{2}+m_{3}^{2}
-m_{1}^{2}),\quad
b_{4}=-m_{3}^{2},\quad
c_{4}=\mu_{s}^{2}. \label{definitionsofaoneatwo}
\end{eqnarray}
Since we assume $m_{3}>m_{1}, m_{2}$, we conclude
\begin{equation}
b_{k}^{2}-4a_{k}c_{k}>0, \quad k=1, 2, 3, 4           \label{deltafromaibici}
\end{equation}
thus there are two zeros for each argument of the logarithms
in Eq.(\ref{integralIonesimplified}). Making use of Eq.(\ref{finalresultofGone})
the final result is calculated
\begin{eqnarray}
I&=&-\frac{\ln(\beta_{0}-\alpha\beta_{1})}{\beta_{1}}
\ln\frac{a_{1}+b_{1}+c_{1}-i\varepsilon}{a_{2}+b_{2}+c_{2}-i\varepsilon}
+\frac{\ln\big[\beta_{0}+(1-\alpha)\beta_{1}\big]}{\beta_{1}}
\ln\frac{a_{3}+b_{3}+c_{3}-i\varepsilon}
{a_{4}+b_{4}+c_{4}-i\varepsilon}\nonumber\\
&+&\frac{1}{\beta_{1}}\sum_{k=1}^{2}
\frac{(-1)^{k-1}}{[\xi_{k}^{(1)}-\xi_{k}^{(1)}]}
\Big\{\Big[2\xi_{k}^{(1)}+\frac{b_{k}}{a_{k}}\Big]
f_{1}\Big(-\alpha\beta_{1},\beta_{0},\xi_{k}^{(+)}\Big)\nonumber\\
&-&\Big[2\xi_{k}^{(2)}+\frac{b_{k}}{a_{k}}\Big]
f_{2}\Big(-\alpha\beta_{1},\beta_{0},\xi_{k}^{(-)}\Big)\Big\}\nonumber\\
&+&\frac{1}{\beta_{1}}\sum_{k=3}^{4}
\frac{(-1)^{k}}{[\xi_{k}^{(1)}-\xi_{k}^{(2)}]}
\Big\{\Big[2\xi_{k}^{(1)}+\frac{b_{k}}{a_{k}}\Big]
f_{1}\Big((1-\alpha)\beta_{1},\beta_{0},\xi_{k}^{(+)}\Big)\nonumber\\
&-&\Big[2\xi_{k}^{(2)}+\frac{b_{k}}{a_{k}}\Big]
f_{2}\Big((1-\alpha)\beta_{1},\beta_{0},\xi_{k}^{(-)}\Big)\Big\}, \label{intIonecompleted}
\end{eqnarray}
where\,$(k=1,2,3,4)$
\begin{eqnarray}
\xi^{(1)}_{k}&=&\xi^{(+)}_{k}+i\varepsilon, \quad\quad
\xi^{(+)}_{k}=\frac{-b_{k}+\sqrt{b_{k}^{2}-4a_{k}c_{k}}}{2a_{k}}, \nonumber\\
\xi^{(2)}_{k}&=&\xi^{(-)}_{k}-i\varepsilon,\quad\quad
\xi^{(-)}_{k}=\frac{-b_{k}-\sqrt{b_{k}^{2}-4a_{k}c_{k}}}{2a_{k}}. \label{definitionofxionexitwo}
\end{eqnarray}

\subsection{ one massive internal line}

\begin{figure}
\begin{center}
\includegraphics[scale=0.75]{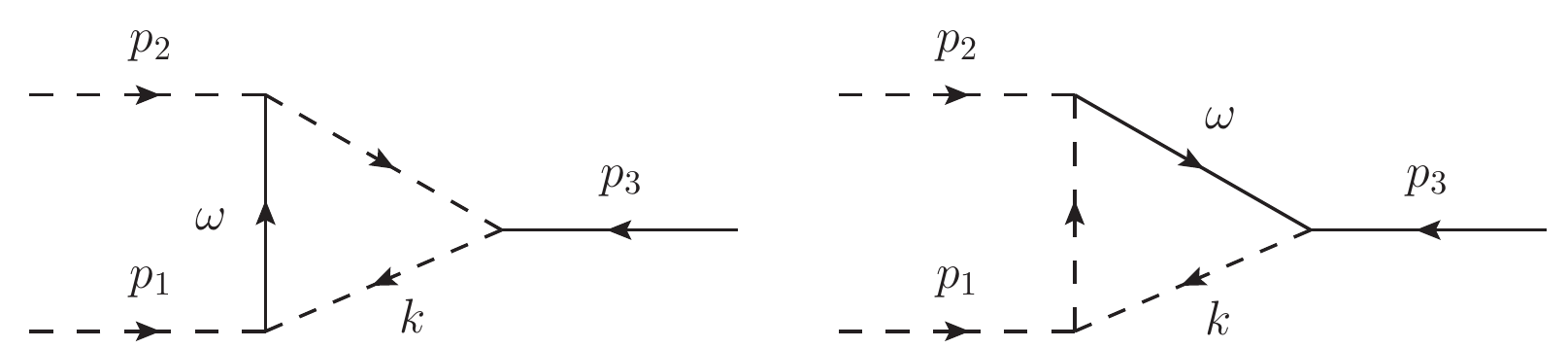}
\caption{Triangle with one massive internal line.}\label{onemassivetriangle}
\end{center}
\end{figure}

(\rm{$\romannumeral 1$})\quad $\omega_{1}=0,
\omega_{2}=\omega, \omega_{3}=0, p_{1}^{2}=p_{2}^{2}=0, p_{3}^{2}=m_{3}^{2}$

The diagram is depicted on the left of fig.\ref{onemassivetriangle},
the amplitude is
\begin{equation}
I=\int_{0}^{1}\D x\int_{0}^{1-x}\D y\frac{1}{W(x,y)}, \label{onemassiveinternallineone}
\end{equation}
where
\begin{equation}
W(x,y)=x^{2}m_{3}^{2}+xym_{3}^{2}
-xm_{3}^{2}+y\omega^{2}+\mu_{s}^{2}-i\varepsilon, \label{onemassivedenomenator}
\end{equation}
The integral over $y$ is easy to evaluate, we obtain
\begin{eqnarray}
I&=&\int_{0}^{1}\D x
\frac{1}{x m_{3}^{2}+\omega^{2}}
\big[\ln\big(1+\frac{\omega^{2}}{\mu_{s}^{2}}
-\frac{\omega^{2}}{\mu_{s}^{2}}x-i\varepsilon\big)
-\ln\big(ax^{2}
+bx+c-i\varepsilon\big)\big], \label{onamssiveyinted}
\end{eqnarray}
with the coefficients
\begin{equation}
a=\frac{m_{3}^{2}}{\mu_{s}^{2}},\quad
b=-\frac{m_{3}^{2}}{\mu_{s}^{2}},\quad c=1,  \label{defabcgroupthreexyz}
\end{equation}
Combining Eq.(\ref{dilogformulatwo}), Eq.(\ref{finalresultofGone}) and Eq.(\ref{finalresultofGtwo}),
we get
\begin{eqnarray}
&&I=\frac{1}{m_{3}^{2}}\big\{\ln(1+\frac{\omega^{2}}{\mu_{s}^{2}}
+\frac{\omega^{4}}{m_{3}^{2}\mu_{s}^{2}})
\ln(1+\frac{m_{3}^{2}}{\omega^{2}})
-{\rm{Li}}_{2}\big[\frac{\omega^{2}(m_{3}^{2}+\omega^{2})}
{\omega^{4}+m_{3}^{2}(\omega^{2}+\mu_{s}^{2})}\big]\nonumber\\
&&+{\rm{Li}}_{2}\big[\frac{\omega^{4}}{\omega^{4}
+m_{3}^{2}(\omega^{2}+\mu_{s}^{2})}\big]\big\}\nonumber\\
&-&\frac{1}{m_{3}^{2}(x_{1}-x_{2})}\big[(2x_{1}-1)
f_{1}(m^{2}_{3},\omega^{2},x^{(+)})
-(2x_{1}-1)
f_{2}(m^{2}_{3},\omega^{2},x^{(-)})\big]. \label{onemassivecaseonefinal}
\end{eqnarray}
where
\begin{eqnarray}
x_{1}&=&x^{(+)}+i\varepsilon,\quad \quad x^{(+)}=\frac{-b+\sqrt{b^{2}-4ac}}{2a},\nonumber\\
x_{2}&=&x^{(-)}-i\varepsilon,\quad \quad x^{(-)}=\frac{-b-\sqrt{b^{2}-4ac}}{2a}, \label{tworootsxyzad}
\end{eqnarray}
for $m_{3}>2\mu_{s}$, and
\begin{eqnarray}
I&=&-\frac{1}{m_{3}^{2}}\big\{{\rm{Li}}_{2}\big[\frac{\omega^{2}(m_{3}^{2}+\omega^{2})}
{\omega^{4}+m_{3}^{2}(\omega^{2}+\mu_{s}^{2})}\big]
+{\rm{Li}}_{2}\big[\frac{\omega^{4}}{\omega^{4}+m_{3}^{2}(\omega^{2}+\mu_{s}^{2})}\big]\big\}\nonumber\\
&+&\frac{2}{m_{3}^{2}}\big\{{\rm{Li}}_{2}
\big[\frac{\omega^{2}}{\sqrt{\omega^{4}+m_{3}^{2}(\omega^{2}+\mu_{s}^{2})}},\theta\big]
-{\rm{Li}}_{2}\big[\frac{\omega^{2}+m_{3}^{2}}
{\sqrt{\omega^{4}+m_{3}^{2}(\omega^{2}+\mu_{s}^{2})}},\theta\big]\big\}, \label{onemassivecasetwodid}
\end{eqnarray}
for $m_{3}<2\mu_{s}$, the parameter $\theta$ is determined by
\begin{equation}
\theta=\arccos\frac{2\omega^{2}+m_{3}^{2}}
{2\sqrt{\omega^{4}+m_{3}^{2}(\omega^{2}+\mu_{s}^{2})}}. \label{onemassivethreepara}
\end{equation}

(\rm{$\romannumeral 2$})\quad $\omega_{1}=\omega_{2}=0, \omega_{3}=\omega,
p_{1}^{2}=p_{2}^{2}=0, p_{3}^{2}=m_{3}^{2}$

The diagram is depicted on the right of fig.\ref{onemassivetriangle},
the corresponding amplitude is
\begin{equation}
I=\int_{0}^{1}\D x\int_{0}^{1-x}\D y\frac{1}{W(x,y)}, \label{onemassivetwo}
\end{equation}
where
\begin{equation}
W(x,y)=x^{2}m_{3}^{2}+xym_{3}^{2}-x(m_{3}^{2}
+\omega^{2})-y\omega^{2}+\omega^{2}+\mu_{s}^{2}-i\varepsilon, \label{onemassivetwofunxy}
\end{equation}
After integral over $y$ is carried out, we arrive at
\begin{eqnarray}
I&=&\int_{0}^{1}\D x\,\frac{\ln\big(ax^{2}+bx+c-i\varepsilon\big)}
{x m_{3}^{2}-\omega^{2}}, \label{onemassivetwoyinted}
\end{eqnarray}
with the coefficients
\begin{equation}
a=\frac{m_{3}^{2}}{\mu_{s}^{2}},
\quad b=-\frac{m_{3}^{2}+\omega^{2}}{\mu_{s}^{2}},\quad
c=\frac{\omega^{2}}{\mu_{s}^{2}}+1. \label{definitionofabcmassivetwo}
\end{equation}
By making use Eq.(\ref{finalresultofGone}) and Eq.(\ref{finalresultofGtwo}),
yielding the following results
\begin{eqnarray}
I&=&-\frac{1}{m_{3}^{2}}\ln\frac{\omega^{2}}{\mu^{2}}\ln(c-i\varepsilon)
-\frac{1}{m_{3}^{2}(x_{1}-x_{2})}\big[(2x_{1}
+\frac{b}{a})f_{1}(m_{3}^{2},-\omega^{2},x^{(+)})\nonumber\\
&-&(2x_{2}+\frac{b}{a})f_{2}(m_{3}^{2},-\omega^{2},x^{(-)})\big],  \label{massiveonecasetwofinalone}
\end{eqnarray}
for $(m_{3}^{2}-\omega^{2})^{2}>4m_{3}^{2}\mu_{s}^{2}$, where the scale $\mu$ is introduced to assure the
argument of logarithm is dimensionless, the two roots are
\begin{eqnarray}
x_{1}&=&x^{(+)}+i\varepsilon,\quad \quad x^{(+)}=\frac{-b+\sqrt{b^{2}-4ac}}{2a},\nonumber\\
x_{2}&=&x^{(-)}-i\varepsilon,\quad \quad x^{(-)}=\frac{-b-\sqrt{b^{2}-4ac}}{2a},\label{tworootsoftwomassivetwo}
\end{eqnarray}
and
\begin{eqnarray}
I&=&\frac{1}{m_{3}^{2}}\ln\big(1-\frac{\omega^{2}}{m_{3}^{2}}
+\frac{\omega^{4}}{m_{3}^{2}\mu_{s}^{2}}\big)
\ln\big(1-\frac{m_{3}^{2}}{\omega^{2}}\big)\nonumber\\
&+&\frac{2}{m_{3}^{2}}\big\{{\rm{Li}}_{2}
\big[-\big(1-\frac{\mu_{s}^{2}}{\omega^{2}}
+\frac{m_{3}^{2}\mu_{s}^{2}}{\omega^{4}}\big)^{-1/2},\theta\big]\nonumber\\
&-&{\rm{Li}}_{2}\big[\big(\frac{m_{3}^{2}}{\omega^{2}}-1\big)\big(1-\frac{\mu_{s}^{2}}{\omega^{2}}
+\frac{m_{3}^{2}\mu_{s}^{2}}{\omega^{4}}\big)^{-1/2},\theta\big]\big\}, \label{twodilogofmassiveon}
\end{eqnarray}
for $(m_{3}^{2}-\omega^{2})^{2}<4m_{3}^{2}\mu_{s}^{2}$, $\theta$ is fixed by
\begin{equation}
\theta=\arccos\frac{m_{3}^{2}
-\omega^{2}}{2\sqrt{\omega^{4}
+\mu_{s}^{2}(m_{3}^{2}-\omega^{2})}}. \label{secondgroupDthetam}
\end{equation}

\subsection{two massive internal lines}

\begin{figure}
\begin{center}
\includegraphics[scale=0.60]{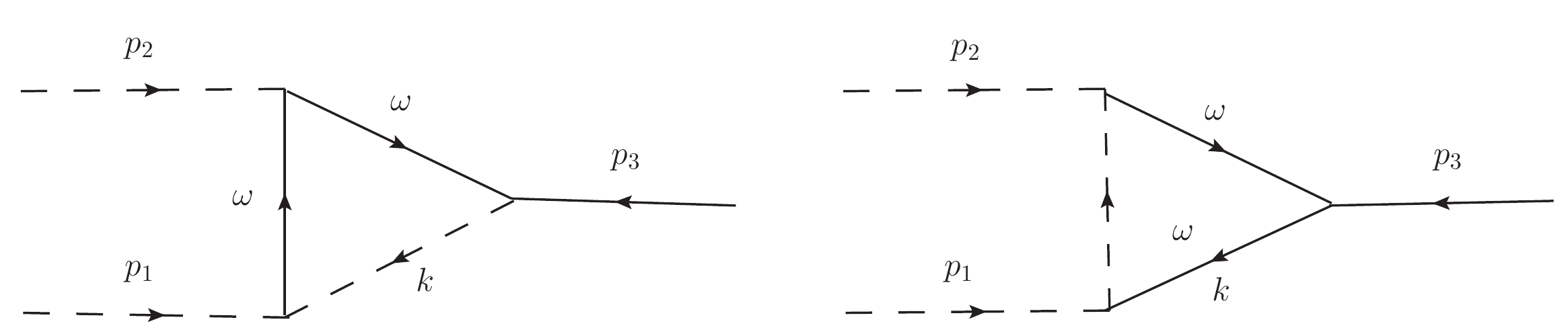}
\caption{Triangle with two massive internal lines.}\label{twomassivetriangle}
\end{center}
\end{figure}

(\rm{$\romannumeral 1$})\quad $\omega_{1}=0, \omega_{2}=\omega_{3}=\omega,
p_{1}^{2}=p_{2}^{2}=0, p_{3}^{2}=m_{3}^{2}$

The diagram is depicted on the left of fig.\ref{twomassivetriangle},
the amplitude is
\begin{equation}
I=\int_{0}^{1}\D x\int_{0}^{1-x}\D y\frac{1}{W(x,y)}, \label{twomassiveprimitive}
\end{equation}
where
\begin{equation}
W(x,y)=x^{2}m_{3}^{2}+xym_{3}^{2}-x(m_{3}^{2}
+\omega^{2})+\omega^{2}+\mu_{s}^{2}-i\varepsilon, \label{twomassivewfuncone}
\end{equation}
After integration over $y$, combining with Eq.(\ref{dilogformulaone}),
Eq.(\ref{finalresultoffirstkindcaseone}) and Eq.(\ref{resultoffirstkindcasetwo}),
obtaining
\begin{eqnarray}
I&=&\frac{1}{m_{3}^{2}}\big[-\int_{0}^{1}\D x
\frac{\ln x}{-x\frac{\omega^{2}}{\mu_{s}^{2}}+1
+\frac{\omega^{2}}{\mu_{s}^{2}}-i\varepsilon}
+\int_{0}^{1}\D x
\frac{\ln(ax^{2}+bx+1-i\varepsilon)}{x}\big]\nonumber\\
&=&-\frac{1}{m_{3}^{2}}\big\{\frac{\mu_{s}^{2}}{(\mu_{s}^{2}
+\omega^{2})}{\rm{Li}}_{2}(1+\frac{\mu_{s}^{2}}{\omega^{2}})
-\frac{1}{2}{\rm{Li}}_{2}(b-a)\nonumber\\
&-&\frac{1}{2}\big(\arcsin\frac{b}{2\sqrt{a}}\big)^{2}
-\arcsin(\frac{b}{2\sqrt{a}})
\arctan\frac{2a-b}{\sqrt{4a-b^{2}}}\nonumber\\
&+&\frac{2a-b}{2\alpha b}
\big[\ln(1+a-b)
+\frac{1}{3}\big(\frac{b^{2}}{2a}
-\frac{\ln(1+a-b)}{\alpha}\big)+\cdots\big]\big\}. \label{twomassivedeltagreaterzero}
\end{eqnarray}
for $b^{2}-4a<0$, where
\begin{equation}
a=\frac{m_{3}^{2}}{\omega^{2}+\mu_{s}^{2}},\quad
b=-\frac{m_{3}^{2}+\omega^{2}}{\omega^{2}+\mu_{s}^{2}}, \quad
\alpha=\frac{4a(a-b)}{b^{2}}, \label{grouponedefinitoofab}
\end{equation}
and
\begin{eqnarray}
I&=&\frac{1}{m_{3}^{2}}
\big\{-\frac{\mu_{s}^{2}}{\mu_{s}^{2}+\omega^{2}}
{\rm{Li}}_{2}(1+\frac{\mu_{s}^{2}}{\omega^{2}})\nonumber\\
&+&\frac{1}{x_{1}-x_{2}}
\big[(\frac{b}{a}-2x_{1}){\rm{Li}}_{2}(\frac{1}{x^{(+)}}-i\varepsilon)
-(\frac{b}{a}-2x_{2}){\rm{Li}}_{2}(\frac{1}{x^{(-)}}+i\varepsilon)\big]\big\}, \label{resultfinaltwomasiveone}
\end{eqnarray}
for $b^{2}-4a>0$, where
\begin{eqnarray}
x_{1}&=&x^{(+)}+i\varepsilon,\quad \quad x^{(+)}=\frac{-b+\sqrt{b^{2}-4ac}}{2a}\nonumber\\
x_{2}&=&x^{(-)}-i\varepsilon,\quad \quad x^{(-)}=\frac{-b-\sqrt{b^{2}-4ac}}{2a} \label{twomassivecaseonetworooots}
\end{eqnarray}

(\rm{$\romannumeral 2$})\quad $\omega_{1}=\omega_{3}=\omega, \omega_{2}=0,
p_{1}^{2}=p_{2}^{2}=0, p_{3}^{2}=m_{3}^{2}$

The diagram is depicted on the right of fig.\ref{twomassivetriangle},
the amplitude is
\begin{equation}
I=\int_{0}^{1}\D x\int_{0}^{1-x}\D y\frac{1}{W(x,y)}, \label{twomassivetwoprimitive}
\end{equation}
where
\begin{equation}
W(x,y)=x^{2}m_{3}^{2}+xym_{3}^{2}-xm_{3}^{2}
-y\omega^{2}+\omega^{2}+\mu_{s}^{2}-i\varepsilon, \label{twomassivecasetwowfun}
\end{equation}
After integration over $y$, we arrive the results
\begin{equation}
I=\int_{0}^{1}\D x
\frac{1}{x m_{3}^{2}-\omega^{2}}
[\ln(x\frac{\omega^{2}}{\mu_{s}^{2}}+1-i\varepsilon)
-\ln(ax^{2}+bx +c-i\varepsilon)],  \label{twomassiveyintegrated}
\end{equation}
with the coefficients
\begin{equation}
a=\frac{m_{3}^{2}}{\mu_{s}^{2}},\quad
b=-\frac{m_{3}^{2}}{\mu_{s}^{2}},\quad
c=1+\frac{\omega^{2}}{\mu_{s}^{2}}, \label{tthirdgroupabc}
\end{equation}
By employing Eq.(\ref{dilogformulatwo}),
Eq.(\ref{resultoffirstkindcasetwo}) and Eq.(\ref{finalresultofGtwo}),
getting
\begin{eqnarray}
I&=&-\frac{1}{\omega^{2}}\big[\ln\frac{m_{3}^{2}+\mu_{s}^{2}}{\mu^{2}}
\ln(1-\frac{\omega^{2}}{m_{3}^{2}})
-{\rm{Li}}_{2}\big(\frac{m_{3}^{2}
-\omega^{2}}{m_{3}^{2}+\mu_{s}^{2}}\big)
+{\rm{Li}}_{2}\big(\frac{m_{3}^{2}}{m_{3}^{2}
+\mu_{s}^{2}}\big)\big]\nonumber\\
&+&\frac{1}{m_{3}^{2}}\ln(1-\frac{m_{3}^{2}}{\omega^{2}})
\ln(1+\frac{\omega^{2}}{\mu_{s}^{2}}-i\varepsilon)\nonumber\\
&-&\frac{1}{m_{3}^{2}(x_{1}-x_{2})}
\big[\big(2x_{1}+\frac{b}{a}\big)f_{1}(m_{3}^{2},-\omega^{2},x^{(+)})
-\big(2x_{2}+\frac{b}{a}\big)f_{2}(m_{3}^{2},-\omega^{2},x^{(-)})\big], \label{twomassivecaseoneresult}
\end{eqnarray}
for $m_{3}>\sqrt{\omega^{2}+\mu_{s}^{2}}$, the two roots are given by
\begin{eqnarray}
x_{1}&=&x^{(+)}+i\varepsilon,\quad x^{(+)}=\frac{-b+\sqrt{b^{2}-4ac}}{2a}\nonumber\\
x_{2}&=& x^{(-)}-i\varepsilon, \quad x^{(-)}=\frac{-b-\sqrt{b^{2}-4ac} }{2a}, \label{tworootsoftwomassive}
\end{eqnarray}
and
\begin{eqnarray}
I&=&-\frac{1}{\omega^{2}}\big[\ln\frac{m_{3}^{2}+\mu_{s}^{2}}{\mu^{2}}
\ln(1-\frac{\omega^{2}}{m_{3}^{2}})
-{\rm{Li}}_{2}\big(\frac{m_{3}^{2}
-\omega^{2}}{m_{3}^{2}+\mu_{s}^{2}}\big)
+{\rm{Li}}_{2}\big(\frac{m_{3}^{2}}{m_{3}^{2}
+\mu_{s}^{2}}\big)\big]\nonumber\\
&+&\frac{1}{m_{3}^{2}}\{\frac{1}{m_{3}^{2}}\ln(1-\frac{m_{3}^{2}}{\omega^{2}})
\ln(1+\frac{\omega^{2}}{\mu_{s}^{2}}-i\varepsilon)
+\ln(1+\frac{\omega^{4}}{m_{3}^{2}\mu_{s}^{2}})
\ln(1-\frac{m_{3}^{2}}{\omega^{2}})\nonumber\\
&+&2\big[{\rm{Li}}_{2}(-\frac{\omega^{2}}{\sqrt{\omega^{4}
+\mu_{s}^{2}m_{3}^{2}}},\theta)
-{\rm{Li}}_{2}(\frac{m_{3}^{2}-\omega^{2}}{\sqrt{\omega^{4}
+\mu_{s}^{2}m_{3}^{2}}},\theta)\big]\},  \label{finalresultoftwomassivecasetwo}
\end{eqnarray}
for $m_{3}<\sqrt{\omega^{2}+\mu_{s}^{2}}$, where
\begin{equation}
\theta=\arccos\big(\frac{m_{3}^{2}-2\omega^{2}}{2\sqrt{\omega^{4}
+\mu_{s}^{2}m_{3}^{2}}}\big). \label{lastgroupDmtheta}
\end{equation}

\section{discussions}
In this paper the infrared divergent one-loop scalar triangle diagrams
with one- two- and three massless internal lines are evaluated through
the loop regularization scheme. Analytic results of each type diagram
are obtained. From these $\mu_{s}$-dependent expressions, the absorptive
and dispersive parts can be extracted. The role played by the sliding
scale $\mu_{s}$ is infrared cutoff. From the primitive expression
Eq.(\ref{intermediateintegrated}), it is obvious that we may extract
different contributions by tuning $\mu_{s}$, i.e., the smaller value
we take, the greater contribution we get since we approach to the
lower energy region. An advantage of the $\mu_{s}$-dependent
of the amplitudes is that we may obtain the desired results which well agree with
experiments by tuning $\mu_{s}$ to some appropriate value.
We should realize that the results obtained
in this way is arbitrary in some sense, a way to fix $\mu_{s}$ should be
suggested in applying loop regularization to evaluate the amplitudes.

In considering the role played by $\mu_{s}$, the most economic way
is that we regard it as the scale of quark confinement in QCD.
Another way determining $\mu_{s}$ is that we may take it as
the scale of factorization at which the light-cone wave function
is defined. The third and feasible way maybe we consider $\mu_{s}$
as an inputting parameter in the evaluation of amplitudes.
If the observables, such as branch ratio, lifetime of many decay channels
are consistent with the experiments for some value of $\mu_{s}$,
then we may take the corresponding $\mu_{s}$ as a reasonable infrared cutoff.
This will be one of our future work in applying and testing of loop regularization.


\begin{appendix}

\section{useful formula}
In this section we list some necessary formula in our evaluation.
The dilogarithm is defined as\cite{lewindilog}
\begin{equation}
{\rm{Li}}_{2}(x)=\sum_{k=1}^{\infty}\frac{x^{k}}{k^{2}}
=\int_{0}^{1}\frac{\ln(1-xt)}{t}\D t, \quad |x|<1\label{definitionofdilog}
\end{equation}
Since there is a branch cut from $1$ to $\infty$, for $\varepsilon \rightarrow 0$
\begin{equation}
{\rm{Li}}_{2}(x+i\varepsilon)
={\rm{Re}}\,{\rm{Li}}_{2}(x)+i\pi\,{\rm{sgn}}(\varepsilon)\Theta(x-1)\ln x,  \label{analyticacontofdilog}
\end{equation}
where $\Theta$ is the step function, the ${\rm{sgn}}(x)$ is
\begin{equation}
{\rm{sgn}}(x)=
\begin{cases}
1  \quad\quad\quad x>0\nonumber\\
-1 \quad\quad x<0
\end{cases}  \label{signfunction}
\end{equation}
Integrals frequently used in this paper are\cite{lewindilog, Devoto:1983tc}
\begin{eqnarray}
\int_{0}^{1}\frac{\ln x}{a+bx}\D x&=&\frac{1}{b}{\rm{Li}}_{2}(-\frac{b}{a}), \label{dilogformulaone}\\
\int_{0}^{1}\frac{\ln (c+ex)}{a+bx}\D x&=&\frac{1}{b}\big\{\ln\frac{bc-ae}{b}\ln\frac{a+b}{a}
-{\rm{Li}}_{2}[\frac{e(a+b)}{ae-bc}]+{\rm{Li}}_{2}(\frac{ae}{ae-bc})\big\},\label{dilogformulatwo} \\
{\rm{Li}}_{2}(r,\theta)&=&{\rm{Re}}\,{\rm{Li}}_{2}(xe^{i\theta})=
-\frac{1}{2}\int_{0}^{r}\frac{\ln(x^{2}-2x\cos\theta+1)}{x}\D x. \label{frequentlyusedformula}
\end{eqnarray}

\section{integral of the first kind}

The integral of the first kind is of the following form
\begin{equation}
I=\int_{0}^{1}\D x\frac{\ln(ax^2-bx+1-i\varepsilon)}{x},
\quad a>0,\quad \varepsilon \rightarrow 0^{+} \label{generallogintegration}
\end{equation}
\begin{itemize}
\item case \rm{$\romannumeral 1$}: $b^{2}-4a<0$
\end{itemize}
In this case the argument of the logarithm is positive-definite thus the $i\varepsilon$ may be safely
dropped. An appropriate way to do the integration turns out to be writing the integral as
\begin{eqnarray}
I&=&\int_{0}^{1}\D x\int_{0}^{1}\D z
\frac{ax-b}{1+zx(ax-b)}\nonumber\\
&=&a\int_{0}^{1}\D z\int_{0}^{1}
\D x\frac{x}{1+zx(ax-b)}
-b\int_{0}^{1}\D z\int_{0}^{1}
\D x\frac{1}{1+zx(ax-b)},    \label{fristparamtering}
\end{eqnarray}
After integral over $x$ of the first term in Eq.(\ref{fristparamtering}) is evaluated,
we get
\begin{eqnarray}
I&=&\frac{1}{2}\int_{0}^{1}\D z\frac{\ln[1+z(a-b)]}{z}
-\frac{b}{2}\int_{0}^{1}\D z\int_{0}^{1}
\D x\frac{1}{1+zx(ax-b))}\nonumber\\
&=&-\frac{1}{2}{\rm{Li}}_{2}(b-a)
-\frac{b}{2}\int_{0}^{1}\D z\int_{0}^{1}
\D x\frac{1}{1+zx(ax-b)}.  \label{variablexintedfirst}
\end{eqnarray}
Now we concentrate on the last integral in Eq.(\ref{variablexintedfirst}).
For later convenience we label the last integral in Eq.(\ref{variablexintedfirst}) as $A$,
the integral over $x$ can be calculated\cite{integraltable2014}
\begin{eqnarray}
A&=&\int_{0}^{1}\D z\int_{0}^{1}
\D x\frac{1}{1+zx(ax-b)}\nonumber\\
&=&\int_{0}^{1}\D z\frac{1}{\sqrt{4az-b^2z^2}}
\big(\arctan\frac{bz}{\sqrt{4az-b^{2}z^{2}}}
+\arctan\frac{2az-bz}{\sqrt{4az-b^{2}z^{2}}}\big), \label{integrationofzA}
\end{eqnarray}
in deriving Eq.(\ref{integrationofzA}) we have using the property that
$\arctan(x)$ is odd
\begin{equation}
\arctan(-x)=-\arctan(x),  \label{oddarctan}
\end{equation}
To proceed we separate the integral into two parts
\begin{eqnarray}
A_{1}&=&\int_{0}^{1}\D z\frac{1}{\sqrt{4az-b^2z^2}}
\arctan\frac{bz}{\sqrt{4az-b^{2}z^{2}}},\nonumber\\
A_{2}&=&\int_{0}^{1}\D z\frac{1}{\sqrt{4az-b^2z^2}}
\arctan\frac{2az-bz}{\sqrt{4az-b^{2}z^{2}}}.  \label{AoneandAtwointegrals}
\end{eqnarray}

We first consider the special case $a=b$. In this case $A$ reduces to
\begin{equation}
A=2\int_{0}^{1}\D z\frac{1}{\sqrt{4az-a^2z^2}}
\arctan\frac{az}{\sqrt{4az-a^{2}z^{2}}},  \label{casespecialaequaltob}
\end{equation}
It is not difficult to show
\begin{equation}
\big[\arcsin\big(b\sqrt{\frac{z}{4a}}\big)\big]'
=\big(\arctan\frac{bz}{\sqrt{4az-b^{2}z^{2}}}\big)'
=\frac{b}{2}\frac{1}{\sqrt{4az-b^{2}z^{2}}}, \label{derivetivesoftwoarcfun}
\end{equation}
then we immediately obtain
\begin{equation}
A=\frac{a}{2}\big(\arctan\frac{a}{\sqrt{4a-a^{2}}}\big)^{2}, \label{specialinted}
\end{equation}
Using the identity
\begin{equation}
\arctan x=\arcsin\frac{x}{\sqrt{1+x^{2}}},  \label{arctantoarcsin}
\end{equation}
we end up with
\begin{equation}
I=-\big(\arcsin\frac{\sqrt{a}}{2}\big)^{2}, \quad\quad  0<a<4. \label{finalresultofspecialcase}
\end{equation}

Now we turn to the general case of $a\neq b$.
By employing Eq.(\ref{derivetivesoftwoarcfun}),
the integral of $A_{1}$ is easy to calculate
\begin{equation}
A_{1}=\frac{1}{b}
\big(\arcsin\frac{b}{\sqrt{4a}}\big)^{2},  \label{resultofgeneralAone}
\end{equation}
The $A_{2}$ integral is
\begin{eqnarray}
A_{2}&=&\frac{2}{b}\arcsin(\frac{b}{2\sqrt{a}})
\arctan\frac{2a-b}{\sqrt{4a-b^{2}}}\nonumber\\
&-&\frac{2(2a-b)}{b\sqrt{a}}\int_{0}^{1}
\frac{\arcsin\big(b\sqrt{\frac{z}{4a}}\,\big)}
{[1+(a-b)z]\sqrt{z}\sqrt{1-\frac{b^{2}}{4a}z}}
\D z,  \label{integrationofatwofirst}
\end{eqnarray}
In order to evaluate the remaining integral in Eq.(\ref{integrationofatwofirst})
we make variable substitution
\begin{eqnarray}
u=b\sqrt{\frac{z}{4a}},  \label{variablesunofz}
\end{eqnarray}
such that the second term in Eq.(\ref{integrationofatwofirst}) cast into
\begin{eqnarray}
B&=&\frac{2(2a-b)}{b^{2}}\int_{0}^{b/(2\sqrt{a})}
\frac{\arcsin u}{(1+\alpha u^{2})\sqrt{1-u^{2}}}\,\D u,\quad \quad
\alpha=\frac{4a(a-b)}{b^{2}},  \label{integralBtranformed}
\end{eqnarray}
Since $0<u<1$ we make the following expansion\cite{handbookabramowitz}
\begin{equation}
\frac{\arcsin u}{\sqrt{1-u^{2}}}
=\sum_{n=0}^{\infty}\frac{2^{n}n!}{(2n+1)!!}u^{2n+1}, \label{axpansionofarcsinandsquare}
\end{equation}
then we get
\begin{eqnarray}
B&=&\frac{2(2a-b)}{b^{2}}
\sum_{n=0}^{\infty}\frac{2^{n}n!}{(2n+1)!!}\int_{0}^{b/(2\sqrt{a})}
\frac{u^{2n+1}}{1+\alpha u^{2}}\,\D u\nonumber\\
&=&\frac{2(2a-b)}{b^{2}}
\big[\frac{1}{2\alpha}\ln\big(1+a-b)+
\frac{b^{2}}{12\alpha a}
-\frac{1}{3\alpha^{2}}\ln(1+a-b)+\cdots\big], \label{integralofBcompleted}
\end{eqnarray}
Combining $A_{1}$ and $A_{2}$, the final result of $I$ in the case $a\neq b$ is
\begin{eqnarray}
I&=&-\frac{1}{2}{\rm{Li}}_{2}(b-a)
-\frac{1}{2}\big(\arcsin\frac{b}{2\sqrt{a}}\big)^{2}
-\arcsin(\frac{b}{2\sqrt{a}})
\arctan\frac{2a-b}{\sqrt{4a-b^{2}}}\nonumber\\
&+&\frac{2a-b}{2\alpha b}
\big\{\ln(1+a-b)
+\frac{1}{3}\big[\frac{b^{2}}{2a}
-\frac{\ln(1+a-b)}{\alpha}\big]+\cdots\big\}.  \label{finalresultoffirstkindcaseone}
\end{eqnarray}

\begin{itemize}
\item case \rm{$\romannumeral 2$}: $b^{2}-4a>0$
\end{itemize}
By employing integration by parts it is easy to get
\begin{equation}
I=-\int_{0}^{1}\frac{(2ax-b)\ln x}{ax^2-bx+1-i\varepsilon}\D x
=-\int_{0}^{1}\frac{(2ax-b)\ln x}{a(x-x_{1})(x-x_{2})}\D x, \label{firstkindcasetwo}
\end{equation}
where
\begin{eqnarray}
x_{1}&=&x^{(+)}+i\varepsilon,\quad \quad
x^{(+)}=\frac{b+\sqrt{b^{2}-4a}}{2a},\nonumber\\
x_{2}&=&x^{(-)}-i\varepsilon,\quad\quad
x^{(-)}=\frac{b-\sqrt{b^{2}-4a}}{2a}, \label{tworootsoffirskindcasetwo}
\end{eqnarray}
By using partial fraction expansion and Eq.(\ref{dilogformulaone}),
we obtain the following result
\begin{eqnarray}
I&=&\frac{1}{x_{1}-x_{2}}(\frac{b}{a}-2x_{1})\int_{0}^{1}\frac{\ln x}{x-x_{1}}\D x
+\frac{1}{x_{1}-x_{2}}(2x_{2}-\frac{b}{a})
\int_{0}^{1}\frac{\ln x}{x-x_{2}}\D x\nonumber\\
&=&\frac{1}{x_{1}-x_{2}}\big\{(\frac{b}{a}-2x_{1})
{\rm{Li}}_{2}[\frac{1}{x^{(+)}}-i\varepsilon]
-(\frac{b}{a}-2x_{2}){\rm{Li}}_{2}
[\frac{1}{x^{(-)}}+i\varepsilon]\big\}.  \label{resultoffirstkindcasetwo}
\end{eqnarray}
Supposing $a=b$, Eq.(\ref{resultoffirstkindcasetwo}) reduces to
\begin{equation}
I=-{\rm{Li}}_{2}[\frac{1}{x^{(+)}}-i\varepsilon]
-{\rm{Li}}_{2}[\frac{1}{x^{(-)}}+i\varepsilon], \label{specialcaseofaequaltob}
\end{equation}
in this case $x^{(+)}$ and $x^{(-)}$ are simplified to
\begin{equation}
x^{(+)}=\frac{1+\sqrt{1-\frac{a}{4}}}{2},\quad \quad
x^{(-)}=\frac{1-\sqrt{1-\frac{a}{4}}}{2}, \label{simplifiedxplusandminus}
\end{equation}
combining Eq.(\ref{analyticacontofdilog}) we obtain
\begin{equation}
I=-\frac{1}{2}\big(\pi+i\ln\frac{1+\sqrt{1-\frac{a}{4}}}
{1-\sqrt{1-\frac{4}{a}}}\big)^{2}.\quad \quad  a>4\label{finalresultaequaltob}
\end{equation}

\section{integral of the second kind}
The general form of the integral of the second kind is
\begin{equation}
G=\int_{0}^{1}\frac{\ln(ax^{2}+bx+c-i\varepsilon)}{Ax+B}\D x,
\quad a>0, \quad \varepsilon \rightarrow 0^{+}  \label{secondkindindef}
\end{equation}
\begin{itemize}
\item case \rm{$\romannumeral 1$}: $a>0,\quad b^{2}-4ac>0$
\end{itemize}
By employing integration by parts we get
\begin{eqnarray}
G_{1}&=&\frac{1}{A}\big[\ln(A+B)\ln(a+b+c-i\varepsilon)-\ln B\ln(c-i\varepsilon)\nonumber\\
&-&\int_{0}^{1}\frac{2ax+b}{ax^{2}+bx+c-i\varepsilon}\ln(Ax+B)\D x\big],  \label{greaterthanzerofirstint}
\end{eqnarray}
Since in this case there are two zeros for the argument of the logarithm,
which reads
\begin{eqnarray}
x_{1}&=&x^{(+)}+i\varepsilon, \quad x^{(+)}=\frac{-b+\sqrt{b^{2}-4ac}}{2a}\nonumber\\
x_{2}&=&x^{(-)}-i\varepsilon, \quad x^{(-)}=\frac{-b-\sqrt{b^{2}-4ac}}{2a}.  \label{tworootsofappen}
\end{eqnarray}
Then by making use the partial fraction expansion on the denominator of Eq.(\ref{greaterthanzerofirstint}),
after the integral in Eq.(\ref{greaterthanzerofirstint}) completed, obtaining
\begin{eqnarray}
G_{1}&=&\frac{1}{A}\big[\ln(A+B)\ln(a+b+c-i\varepsilon)-\ln B\ln(c-i\varepsilon)\big]\nonumber\\
&-&\frac{1}{A(x_{1}-x_{2})}
\big[\big(2x_{1}+\frac{b}{a}\big)\int_{0}^{1}\frac{\ln(Ax+B)}{x-x_{1}}\D x
-\big(2x_{2}+\frac{b}{a}\big)\int_{0}^{1}\frac{\ln(Ax+B)}{x-x_{2}}\D x\big]\nonumber\\
&=&\frac{1}{A}[\ln(A+B)\ln(a+b+c-i\varepsilon)-\ln B\ln(c-i\varepsilon)]\nonumber\\
&-&\frac{1}{A(x_{1}-x_{2})}
\big[\big(2x_{1}+\frac{b}{a}\big)f_{1}(A,B,x^{(+)})
-\big(2x_{2}+\frac{b}{a}\big)f_{2}(A,B,x^{(-)})\big],  \label{finalresultofGone}
\end{eqnarray}
where the analytic continuation of the logarithm
\begin{equation}
\ln(-x+i\varepsilon)=\ln x+i\pi\,{\rm{sgn}}(\varepsilon), \quad x>0  \label{continuationoflog}
\end{equation}
has been used, the two functions $f_{1}$ and $f_{2}$ are defined as
\begin{eqnarray}
f_{1}(A, B, x)&=&\ln(Ax+B+i\varepsilon)[\ln(\frac{1}{x}-1)+i\pi]\nonumber\\
&-&{\rm{Li}}_{2}\big\{\frac{x-1}{x+\frac{B}{A}}
+i[1+{\rm{sgn}}(x+\frac{B}{A})]\varepsilon\big\}\nonumber\\
&+&{\rm{Li}}_{2}\big\{\frac{x}{x+\frac{B}{A}}
-i[1-{\rm{sgn}}(x+\frac{B}{A})]\varepsilon\big\},\nonumber\\
f_{2}(A, B, x)&=&\ln(Ax+B-i\varepsilon)[\ln(\frac{1}{x}-1)-i\pi]\nonumber\\
&-&{\rm{Li}}_{2}\big\{\frac{x-1}{x+\frac{B}{A}}
-i[1+{\rm{sgn}}(x+\frac{B}{A})]\varepsilon\big\}\nonumber\\
&+&{\rm{Li}}_{2}\big\{\frac{x}{x+\frac{B}{A}}
+i[1-{\rm{sgn}}(x+\frac{B}{A})]\varepsilon\big\}, \label{functionfoneandtwodef}
\end{eqnarray}

\begin{itemize}
\item case \rm{$\romannumeral 2$}: $a>0,\quad b^{2}-4ac<0$
\end{itemize}
\begin{equation}
G_{2}=\int_{0}^{1}\frac{\ln(ax^{2}
+bx+c-i\varepsilon)}{Ax+B}\D x,  \label{secondkinddeatalesszero}
\end{equation}
Since the argument is positive-definite,
the $i\varepsilon$ term can be dropped.
We make the following variable substitution
\begin{equation}
Ax+B=mu,   \label{lineartransformtou}
\end{equation}
After some evaluation, combining Eq.(\ref{frequentlyusedformula}), the result reads
\begin{equation}
G_{2}=\frac{1}{A}\big\{\ln\frac{D}{A^{2}}\ln(1+\frac{A}{B})
+\big[{\rm{Li}}_{2}(\frac{B}{m},\theta)
-{\rm{Li}}_{2}(\frac{A+B}{m},\theta)\big]\big\}, \label{finalresultofGtwo}
\end{equation}
where
\begin{eqnarray}
D&=&aB^{2}-ABb+cA^{2},\quad m=\sqrt{\frac{D}{a}},\nonumber\\
\theta&=&\arccos\frac{m(2aB-Ab)}{2D}. \label{definitionofDmtheta}
\end{eqnarray}

\end{appendix}

\end{document}